\documentclass{prop2015}
\usepackage[english]{babel}
\global\arraycolsep=2pt
\usepackage{stmaryrd}
\usepackage{amsmath}
\usepackage{amssymb}
\usepackage{graphicx}
\usepackage{textcomp}
\usepackage{calrsfs}
\usepackage{yfonts}
\usepackage{bm}

\newcommand{\ben}{\begin{equation*}}                                                 
\newcommand{\een}{\end{equation*}}
\newcommand{\bean}{\begin{eqnarray*}}                                                
\newcommand{\eean}{\end{eqnarray*}}

\newcommand{\balpha}{\mbox{\boldmath{$\alpha$}}}
\newcommand{\bbeta}{\mbox{\boldmath{$\beta$}}}
\newcommand{\bone}{\mbox{\boldmath{$1$}}}
\newcommand{\bepsilon}{\mbox{\boldmath{$\epsilon$}}}
\newcommand{\nn}{\nonumber}
\newcommand{\be}{\begin{equation}}                                                   
\newcommand{\ee}{\end{equation}}
\newcommand{\bea}{\begin{eqnarray}}                                                  
\newcommand{\eea}{\end{eqnarray}}
\DeclareMathOperator{\Tr}{Tr}

\title{Negative Entropies in Casimir and Casimir-Polder Interactions}
\author[K. Milton]{Kimball A. Milton\inst{1,}\footnote{Corresponding 
author\quad E-mail:~\textsf{kmilton@ou.edu}}}
\author[Y. Li]{Li Yang\inst{1}}
\author[P. Kalauni]{Pushpa Kalauni\inst{1}}
\author[P. Parashar]{Prachi Parashar\inst{1,2}}
\author[R. Gu\'erout]{Romain Gu\'erout\inst{3}}
\author[G.-L. Ingold]{Gert-Ludwig Ingold\inst{4}}
\author[A. Lambrecht]{Astrid Lambrecht\inst{3}}
\author[S. Reynaud]{Serge Reynaud\inst{3}}
\address[1]{H. L. Dodge Department of Physics and Astronomy,
University of Oklahoma, Norman, OK 73019 USA}
\address[2]{Department of Physics, Southern Illinois University--Carbondale,
Carbondale, IL 62091 USA}
\address[3]{Laboratoire Kastler Brossel, CNRS, ENS, UPMC, Case 74, F-75252,
Paris, France}
\address[4]{Institut f\"ur Physik, Universit\"at Augsburg, 
Universit\"atsstra\ss e 1, D-86135, Germany}
\shortauthors{K. Milton et al.}
\begin{abstract}
 It has been increasingly becoming clear that Casimir and Casimir-Polder 
entropies may be negative in certain regions of temperature and separation.  
In fact, the occurrence of negative entropy seems to be a nearly ubiquitous 
phenomenon.  This is most highlighted in the quantum vacuum interaction of a 
nanoparticle with a conducting plate or between two nanoparticles. 
It has been argued that this phenomenon 
does not violate physical intuition, since the total entropy, including the 
self-entropies of the plate and the nanoparticle, should be positive.  New 
calculations, in fact, seem to bear this out at least in certain cases.
\end{abstract}
\shortabstract
\begin{document}
\maketitle

\section{The thermal Casimir puzzle}
For 15 years there has been a controversy surrounding entropy in the
Casimir effect.  This is most famously centered around the issue of how to
describe
a real metal, in particular, the permittivity $\varepsilon(\omega)$
at zero frequency \cite{bs}. The latter determines the low-temperature
and high temperature corrections to the free energy, and hence to the
entropy.  The issue involves  how $\omega^2\varepsilon(\omega)$ 
behaves as $\omega\to 0$.  Dissipation or finite conductivity
implies this vanishes; this leads
to a linear temperature dependence at low temperature, and a reduction
of the high-temperature force. 
Most experiments \cite{decca,mohideen,bimonte-decca}, 
but not all \cite{sushkov}, 
favor the nondissipative plasma model!  For an overview of the status of
both theory and experiment, see Ref.~\cite{dalvit}.

 The Drude model, and general thermodynamic and electrodynamic
arguments, suggest that the transverse electric (TE) reflection coefficient
at zero frequency for a good, but imperfect, metal plate should vanish.
Careful calculations for lossy parallel plates
show that at very low temperature the free energy
approaches a constant
 quadratically in the temperature, thus forcing the entropy
to vanish at zero temperature \cite{njp}. Thus, there is no violation of 
the third law of thermodynamics.  However, there would persist a region at low
temperature in which the entropy would be negative.  This was not thought
to be a problem, since the interaction 
Casimir free energy does not describe the
entire system of the Casimir apparatus, whose total entropy must necessarily
be positive.  The physical basis for the negative entropy region
remains somewhat
mysterious.  Here we address negative entropy arising from geometry.
The interplay of geometry and material properties is further explored in 
Ref.~\cite{ingold}.

\begin{figure}
  \includegraphics[width=\columnwidth]{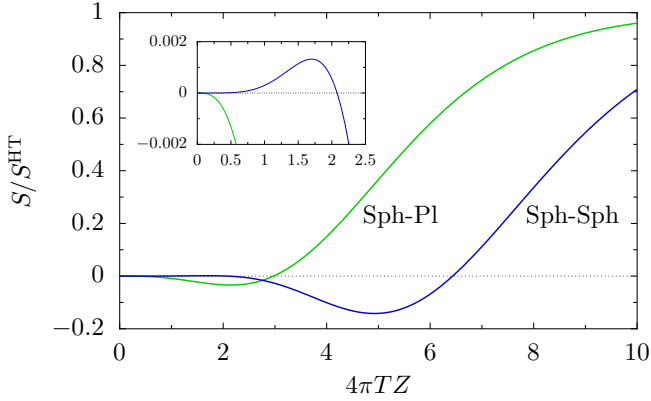}%
  \caption{\label{fig1}\col 
The entropy of interaction between a sphere and a plane (Sph-Pl) and
between two spheres (Sph-Sph) normalized with respect to the respective
high-temperature values is displayed as a function of the product of
distance $Z$ and temperature $T$. The entropy has been evaluated
within the dipole and single-scattering approximation \cite{ingold1}. 
The inset shows the behavior of the
entropy for small $TZ$. We call the negative entropy region perturbative
for the sphere-plane configuration and nonperturbative for the sphere-
sphere configuration.}
%
\end{figure}

For some time it has been known that negative entropy regions can emerge
geometrically.  For example, when a perfectly conducting sphere is near a
perfectly conducting plate, the entropy at room temperature can turn negative,
with enhancement of the effect occurring for smaller spheres \cite{canaguier}.
The occurrence of negative interaction entropies for a small sphere  
in front of a plane or another sphere is illustrated in Fig.~\ref{fig1} within
  the dipole and single-scattering approximation.
Since the negative entropy region is enhanced by making the sphere small,
this suggests that the phenomena be explored by considering the dipole 
approximation only.  That is, we will examine point particles, atoms or
nanoparticles, characterized by electric and magnetic polarizabilities.

Henceforth, we will use natural units with $\hbar=c=k_B=1$.

\section{CP free energy--multiple-scattering approach}
\begin{figure}
\includegraphics[width=\columnwidth]{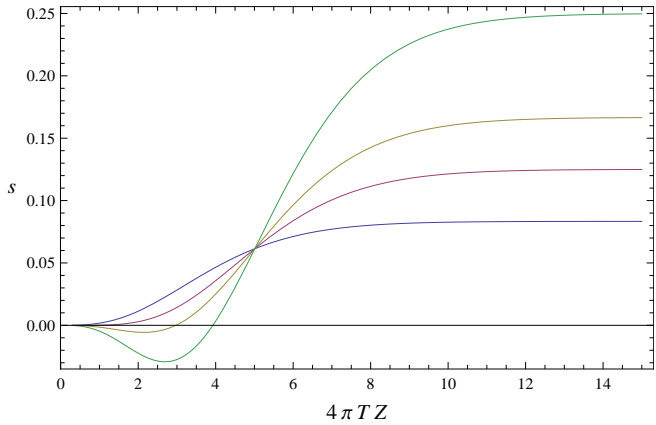}
\caption{\label{fig2} Scaled entropy $s$, defined in Eq.~(\ref{snp}),
between a purely electrically polarizable nanoparticle
and a conducting plate, as a function of the product of the temperature times
the distance from the plate.  The different curves, bottom to top for large 
$ZT$ are for $\gamma= 0$ (blue), $1/2$ (red), $1$ (yellow), $2$ (green).}
\end{figure}

This section is based on Ref.~\cite{milton}. 
The general formula for the free energy between two objects is
\bea
F_{12}&=&\frac12 \Tr \ln(\bm{1}-\bm{\Gamma}_0\mathbf{T}_1^E\bm{\Gamma}_0
\mathbf{T}_2^E)+\frac12
\Tr \ln(\bm{1}-\bm{\Gamma}_0\mathbf{T}_1^M\bm{\Gamma}_0
\mathbf{T}_2^M)\nn\\&&\quad
\mbox{}-\frac12\Tr\ln(\bm{1}+\bm{\Phi}_0\mathbf{T}^E\bm{\Phi}_0
\mathbf{T}^M)\,,\label{ms}
\eea
where the E-M interference term does not separate the contributions from the
two bodies \cite{vac}. 
Here $\bm{\Gamma}_0$ is the free electromagnetic Green's dyadic, and
$\bm{\Phi}_0=-\frac1\zeta\bm{\nabla}\times\bm{\Gamma}_0$ in terms of the
imaginary frequency $\zeta$.
We initially consider the interaction between a
perfectly conducting plate and an electrically polarizable ``atom.'' 
Here appearing is the purely electric 
scattering matrix  for a perfectly conducting  plate, defined from
\be \bm{\Gamma}_0\mathbf{T}_p\bm{\Gamma}_0=\bm{\Gamma-\Gamma}_0,\ee
where the well-known image construction gives
\be
(\bm{\Gamma-\Gamma}_0)(\mathbf{r,r'})=-\bm{\Gamma}_0(\mathbf{r},\mathbf{r}'
-2\mathbf{\hat z}z')\cdot(\bm{1}-2\mathbf{\hat z\hat z})\, .
\ee
For the atom the scattering matrices are identified with the potentials,
\be
\mathbf{T}^E_a=\mathbf{V}^E_a=4\pi \bm{\alpha}\delta
(\mathbf{r-R}),\,
\mathbf{T}^M_a=\mathbf{V}^M_a=4\pi \bm{\beta}\delta
(\mathbf{r-R})\,,\label{tatom}
\ee
and because we can regard this interaction as weak, we can expand the 
logarithms in Eq.~(\ref{ms}) and keep only the first term, the 
single-scattering
approximation.  If we assume the principal axis of the atom aligns
with the direction normal to the plate, and the atom is symmetric around that 
axis, the electric polarizability is
\be
\bm{\alpha}=\mbox{diag}(\alpha_\perp,\alpha_\perp,\alpha_z)\, ,\quad
\gamma=\alpha_\perp/\alpha_z\, ,
\ee which gives the definition of the anisotropy parameter $\gamma$.
In this way we easily obtain the free energy
\be F^E_{ap}=-\frac{3\alpha_z}{8\pi Z^4}f(\gamma,y)\, ,\ee
where we have pulled out the ordinary Casimir-Polder (CP) energy \cite{cp}.
Here,  the reduced free energy is
\be
f(\gamma,y)=\frac{y}6[(1+\gamma)(1-y\partial_y)+\gamma y^2
\partial^2_y]\frac12\coth\frac{y}2
\ee (the normalization is chosen so that $f(1,0)=1$), 
where $y=4\pi Z T$, $Z$ being the separation between the atom and the plate.
The corresponding entropy is
\be
S_{np}^E=-\frac{\partial}{\partial T}F_{np}^E
=\frac{3\alpha_z}{2 Z^3}\frac\partial{\partial y}f(\gamma,y)
=\frac{3\alpha_z}{2 Z^3}s(\gamma,y)\, .\label{snp}
\ee
For large $y$ this entropy approaches a constant,
\be
s(\gamma,y)\sim \frac1{12}(1+\gamma)\, ,\quad y\to\infty\, ,
\ee
while for small $y$,
\be
s(\gamma,y)\sim \frac1{540}(1-2\gamma)y^3+O(y^5)\,.\label{ssmally}
\ee
The entropy vanishes at $T=0$, and then starts off negative for small $y$ when
$\gamma>1/2$.  In particular, even for an isotropic, solely electrically
polarizable, nanoparticle, where $\gamma=1$, the entropy is negative
for a certain region in $y$, as discovered in Ref.~\cite{bezerra}.
The behavior of the entropy with $\gamma$ is illustrated in  Fig.~\ref{fig2}.
For an isotropic nanoparticle,
the negative entropy region occurs for $4\pi Z T<2.97169$,
or at temperature $300$ K, for distances less than 2 $\mu$m.

\begin{figure} 
\includegraphics[width=\columnwidth]{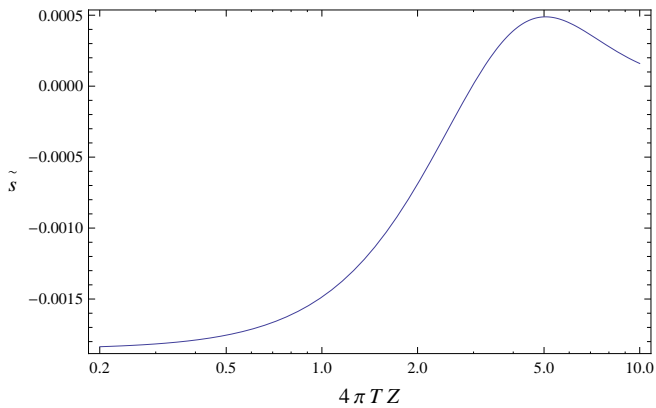}
\caption{\label{fig1a} Rescaled entropy $\tilde s$ plotted as a function of $4\pi ZT$ for $\gamma=1$.}
\end{figure}

Most Casimir experiments are performed at room temperature.  Therefore,
it might be better to present the entropy in the form 
\be
S_{np}^E=\frac{3\alpha_z}2(4\pi T)^3\tilde{s}(\gamma,y)\, ,\quad
\tilde{s}(\gamma,y)=y^{-3} s(\gamma,y)\, ,
\ee
which in view of Eq.~(\ref{ssmally}) makes explicit that the entropy tends to 
a finite value as $Z\to0$.  This version of the entropy for the 
isotropic case is plotted in  
Fig.~\ref{fig1a}.  

\begin{figure}
\includegraphics[width=\columnwidth]{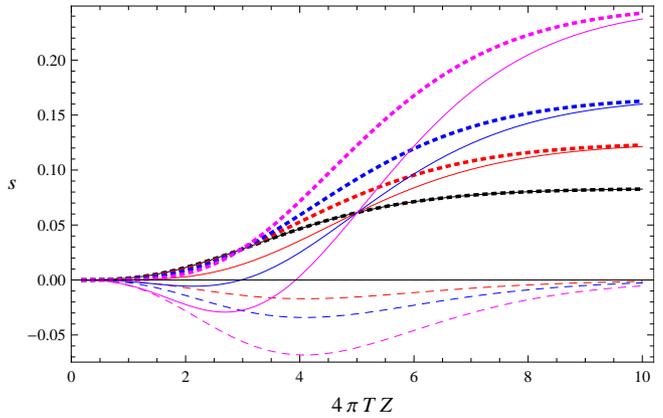}
\caption{\label{fig3} The entropy between an electrically polarizable 
atom or nanoparticle
and a conducting wall. The solid curves are the total entropy, the
short-dashed curves are for the TM plate contribution, and the long-dashed
curves are for TE.
Referring to the ordering for large $TZ$, the inner set of curves (black) 
is for $\gamma=0$, the next set (red) is for $\gamma=1/2$, where the negative
total entropy region starts to appear, the third set 
(blue) is for $\gamma=1$, and the outer set (magenta) is for $\gamma=2$.}
\end{figure}

\begin{figure}
\includegraphics[width=\columnwidth]{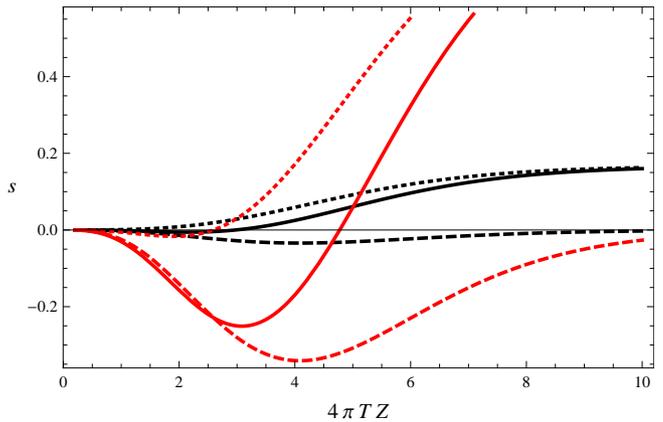}
\caption{\label{fig4} 
The inner set of curves
(black) are for $\gamma=1$, and the outer curves (red) for  $\gamma=10$.
Again, the solid curves show the total entropy $s$, the long-dashed
curves show the TE contribution  $s_E$, and the short-dashed
curves the TM contribution $s_H$.}
\end{figure}

We can also break up the electric response of the plate into TE and TM 
components, which we denote by E and H respectively.
The TE contribution is always negative, while the TM is mostly positive, 
as shown in Fig.~\ref{fig3}.
For sufficient anisotropy, we see in Fig.~\ref{fig4} that 
 even for a solely electrically
polarizable nanoparticle $S_H$ can turn negative, which occurs for $\gamma>2$.
This sign reversal is seen perturbatively, because
\be
s_H(\gamma,y)\sim\frac{y^3}{540}\left(1-\frac{\gamma}2\right)\, ,\quad y\ll1\,.
\ee

The magnetic polarizability contribution from the atom 
is obtained from the electric
polarizability entropy by the replacement $\balpha\to-\bbeta$,
as demonstrated in Ref.~\cite{milton}.
This simple relation between the electric and magnetic polarizability 
contributions was noted earlier in Ref.~\cite{bimonte}. 
In particular, for the interesting case of a conducting sphere of radius
$a$, where
\be
\alpha=-2\beta=a^3\, ,\ee
the previous
results apply, except multiplied by a  factor of 3/2.  In that case,
the limiting value of the entropy is
\be
S(T)\sim -\frac4{15}(\pi a T)^3\, ,\quad 4\pi Z T\ll1\, .\label{mininent}
\ee

\section{CP interaction of 2  nanoparticles}
Let us now consider two nanoparticles, one located at the origin and one at
$\mathbf{R}=(0,0,Z)$.
Let the nanoparticles have both static electric and magnetic polarizabilities
$\bm{\alpha}_i$, $\bm{\beta}_i$, $i=1,2$.  We will again suppose the
nanoparticles
to be anisotropic, but, for simplicity, having their principal axes aligned
with the direction connecting the two nanoparticles, and symmetric around
that axis:
\be
\bm{\alpha}_i=\mbox{diag}(\alpha^i_\perp,\alpha^i_\perp,\alpha_z^i)\, ,\quad
\bm{\beta}_i=\mbox{diag}(\beta^i_\perp,\beta^i_\perp,\beta_z^i)\,.
\ee
The methodology is very similar to that explained in the particle/plate
discussion.    Again, the details are given in Ref.~\cite{milton}.

\begin{figure}
\includegraphics[width=\columnwidth]{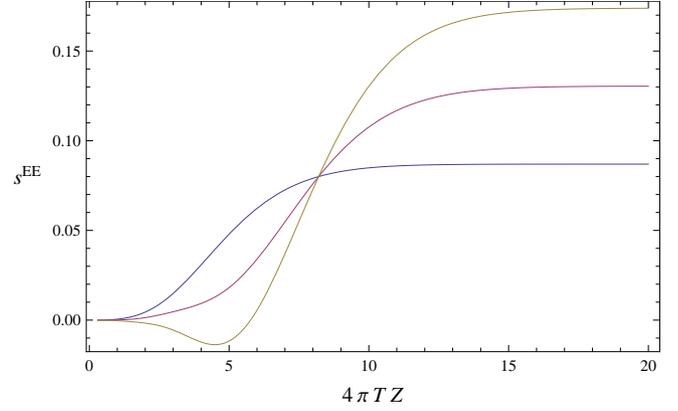}
\caption{\label{figiv} The entropy $s^{EE}(\gamma,y)$
for two anisotropic purely electrically
polarizable nanoparticles
with separation $Z$ and temperature $T$.  When $\gamma=
\gamma_1\gamma_2>1$ the entropy can be negative. The curves, bottom to top 
for large $ZT$ are for $\gamma=0$ (blue), $\gamma= 1$ (red), $\gamma= 2$ 
(yellow).}
\end{figure}

For pure E nanoparticles the interaction entropy is expressed as a 
derivative of the free energy,
\be
S^{EE}=\frac{23\alpha_z^1\alpha_z^2}{Z^6}s^{EE}(\gamma,y)\,,
\quad s^{EE}(\gamma,y)=\frac\partial{\partial y}f(\gamma,y)\, .\label{see}
\ee
Here $\gamma=\gamma_1\gamma_2$ is the product of the anisotropies of the two
atoms.  The asymptotic limits are
\begin{subequations}  
\bea
s^{EE}(\gamma,y)&\sim& \frac{2+\gamma}{23}\,, \quad y\gg1\,,\\
s^{EE}(\gamma,y)&\sim&\frac1{2070}(1-\gamma)y^3\,,\quad y\ll1\,,
\eea
\end{subequations} 
so even in the pure electric case there is a region of negative entropy for
$\gamma>1$, illustrated in Fig.~\ref{figiv}. 
The coupling of two magnetic polarizabilities is given by precisely the same
formulas, except for the replacement $\bm{\alpha}\to\bm{\beta}$.

\begin{figure}
\includegraphics[width=\columnwidth]{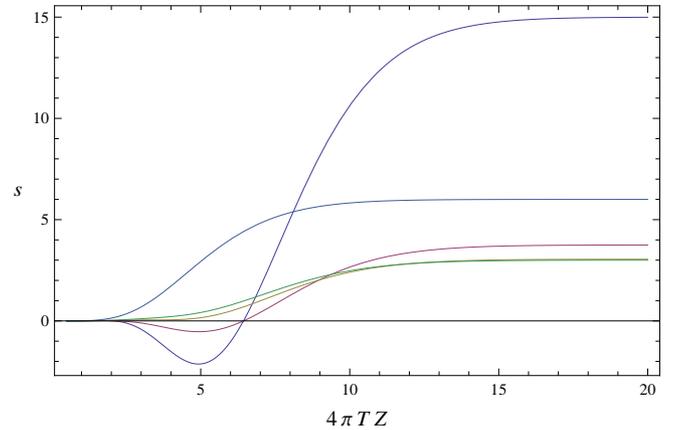}
\caption{\label{fig5} Entropy of two identical isotropic atoms
($\gamma_\alpha=\gamma_\beta=1$) for different values of $r=\beta/\alpha$.  
Starting from highest to lowest curves on the left, the entropy is given for
$r=1$ (purple),  0 (green), $-1/8$ (yellow),$-1/2$ (red),  $-2$ (blue).
What is plotted in this and the following figures is 
$s$, where the entropy is $S=[(\alpha_z^1)^2/Z^6]s$.}
\end{figure}

For atoms possessing both electric and magnetic polarizabilities,
there is also an EM cross term, which comes from the last term in 
Eq.~(\ref{ms}),
\be
S^{EM}=-\frac7{Z^6}(\alpha_\perp^1\beta_\perp^2+\beta_\perp^1\alpha_\perp^2)
s^{EM}\,.
\ee
This involves only the transverse polarizabilities of the atoms; it is always
negative, and also vanishes rapidly for small $y$:
\be
s^{EM}\sim -\frac{y^5}{7056},\quad y\ll1\,.
\ee
Figure \ref{fig5} shows the entropy of identical isotropic atoms, for different
values of the electric and magnetic polarizabilities.

\begin{figure}
\includegraphics[width=\columnwidth]{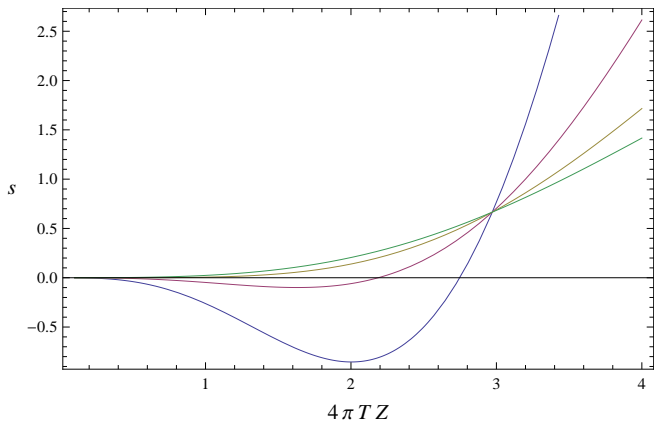}
\caption{\label{fig6} Here the identical nanoparticles
have equal electric and magnetic
polarizabilities, and equal anisotropies. $\gamma=0$ (green), 1 (yellow),
2 (red), 4 (blue), respectively, top to bottom on the left.}
\end{figure}

Now we consider the atoms as having equal polarizabilities 
and equal anisotropies.  Again, as seen perturbatively, the boundary value
for negative entropy is $\gamma=1$, as seen in Fig.~\ref{fig6}.

\begin{figure}
\includegraphics[width=\columnwidth]{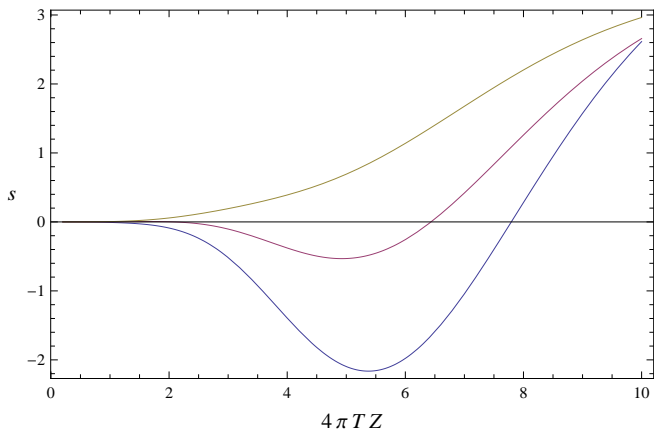}
\caption{\label{fig7}The case of two identical  conducting spheres where 
$\alpha_z=-2\beta_z$,
with electrical  isotropy, but magnetic anisotropy $\gamma_\beta=0$ (yellow),
1 (red), 2 (blue), reading from top to bottom.}
\end{figure}

Particularly interesting is the case of two identical perfectly
 conducting spheres, where $\alpha=-2\beta$, illustrated in Fig.~\ref{fig7}.

\begin{figure}
\includegraphics[width=\columnwidth]{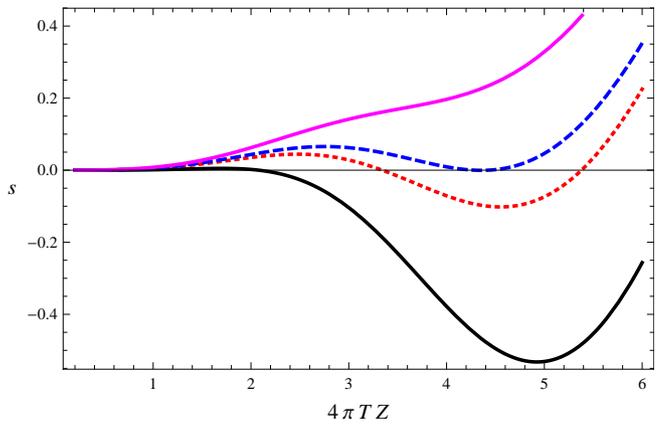}
\caption{\label{fig8} Two identical
nanoparticles  with $\beta_z=-\alpha_z/2$, appropriate for a
conducting sphere, isotropic magnetically,
$\gamma_\alpha= 0.6$ (magenta), 0.743 (dashed blue), 0.8 (short dashed red),
1 (black).}
\end{figure}

In many cases we see instances of nonperturbative negative entropy, by which
we mean that for small $ZT$ the entropy is initially positive, but turns 
negative for larger values of $ZT$, after which it becomes positive again.  
Examples are shown in Figs.~\ref{fig1} and \ref{fig8}.

\begin{figure}
\includegraphics[width=\columnwidth]{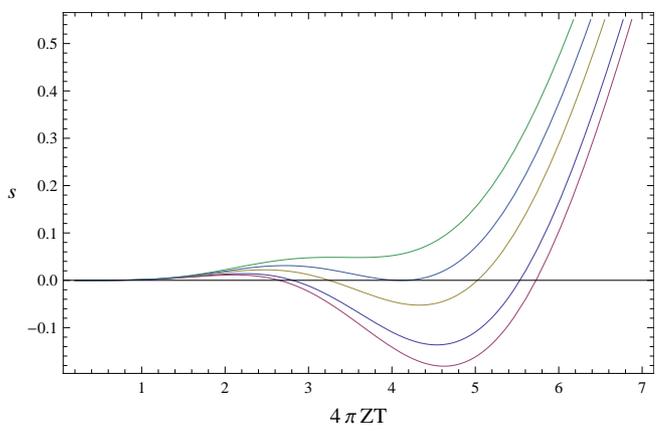}
\caption{\label{fig9}  Interaction entropy between a  perfectly
conducting nanoparticle,
 $\bm{\beta}_1=-\frac12\bm{\alpha}_1$, and a Drude nanoparticle
with the same electric polarizability and no magnetic polarizability,        
with $\bm{\alpha}_2=\bm{\alpha}_1$, $\bm{\beta}_2=0$. Now it is assumed that
the nanoparticles are electrically isotropic, $\gamma_\alpha=1$.
The dependence on the magnetic         
anisotropy of the first nanoparticle is shown.                                 
Reading from top to bottom the magnetic anisotropies are
$\gamma_{\beta 1}=0.5$ (green), 0.66 (purple),  0.8 (yellow),  1 (blue),
1.1 (red).}
\end{figure}

A Drude-model nanoparticle is characterized by having no magnetic 
polarizability.
The interaction between two such isotropic
nanoparticles does not exhibit repulsion, although, as already seen
in Fig.~\ref{figiv}, they would experience negative entropy for $\gamma>1$.
However,
the interaction between a perfectly conducting nanoparticle and Drude 
nanoparticle is more interesting, as seen in Fig.~\ref{fig9}, for equal
electric polarizabilities.

We summarize our findings in the table \ref{tab}.
\begin{table}
 \begin{proptabular}{lr}{\label{tab} 
Summary of circumstances under which negative entropy appears in the
interaction between nanoparticles or between a nanoparticle and a plate.}%
Two nanoparticles
or particle/plate& Negative entropy?\\
E/E& $S<0$ occurs for $\gamma_\alpha>1$\\
E/M &$S<0$ always\\
PC/PC&$S<0$ for $\gamma_\alpha>0.74$ or $\gamma_\beta>0.54$\\
PC/D&$S<0$ for $\gamma_\alpha>0.91$ or $\gamma_\beta>0.66$ \\
E/TE plate&$S<0$ always\\
E/TM plate&$S<0$ for $\gamma_\alpha>2$\\
E/PC or D plate&$S<0$ for $\gamma_\alpha >1/2$\\
 \end{proptabular}
\end{table}

\section{Do self-entropies resolve puzzle?}
We have suggested that positive self-entropies from the bodies nullify this
negative interaction entropy.  But is this possible?
We have considered, in idealized models, the self-entropies of plates and
nanospheres.  A complete account of this work will appear elsewhere \cite{li}.

As a simple model we considered an anisotropic $\delta$-function plate
described by a permittivity \cite{shajesh}
\be
\bepsilon-\bone=\mbox{diag}(\lambda,\lambda,0)\delta(z)\, .
\ee
 The regulated free energy of the plate per unit area $A$ is given by 
\bea
\frac{F}A&=&
-\frac{T}{4\pi}\sum_{m=-\infty}^\infty {\rm e}^{i\zeta_m\tau}\int_0^\infty
{\rm d}k\,k \,J_0(k\delta)\nn\\
&&\qquad\times\left[\ln\frac2{2+\lambda\kappa_m}+\ln \frac{2\kappa_m}
{2\kappa_m+\lambda\zeta_m^2}\right]\, .
\eea  Here the two terms are the TM and the TE modes, respectively, and
$\kappa_m=                   
\sqrt{k^2+\zeta_m^2}$, $\zeta_m=2\pi m T$ being the Matsubara frequency.
This is regulated by point splitting: in (Euclidean) time by a parameter 
$\tau$, and in transverse space by a two-dimensional vector $\bm{\delta}$.

\begin{figure}
\includegraphics[width=\columnwidth]{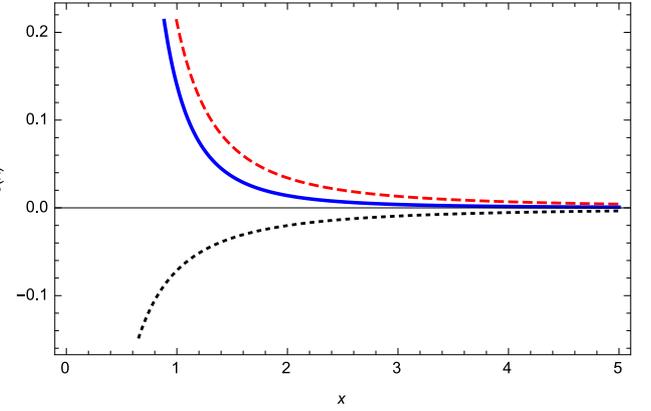}
\caption{\label{pentropy} Total, TE, and TM entropy of a $\delta$-function
plate, shown by solid, dotted, and dashed line. Here the reduced
entropy $s$ is plotted, defined by Eq.~(\ref{splate}).}
\end{figure}

As in realistic materials, it is essential to include dispersion.  Here
we take the plasma model:
$ \lambda=\frac{\lambda_0}{\zeta_m^2}.$  Then the entropy is finite
({\it not the free energy}) and we find the following results for the entropy
per unit area,
\be
\frac{S}A=-\frac\partial{\partial T}\frac{F}A=\frac{\lambda_0^2}{16\pi}s,
\quad x=\frac{\lambda_0}{4\pi T}\, .\label{splate}
\ee
The TE self-entropy is always negative.  The limiting values for low and
high temperatures are
\begin{subequations}
\bea
s^{\rm TE}(x)&\sim& -\frac{3\zeta(3)}{4\pi^2x^2}\,,\quad x\gg1\,,\\
s^{\rm TE}(x)&\sim&-\frac1{3x}+\frac34-\frac12\ln 2\pi x\, ,\quad x\ll1\,.
\eea
 The TM self-entropy is always positive.
The limiting values for low and high temperature is this case are
\bea
s^{\rm TM}(x)&\sim&\frac{3\zeta(3)}{4\pi^2}\frac1{x^2}+\frac1{15}\frac1{x^3}
+\frac{15\zeta(5)}{4\pi^4}\frac1{x^4}\,,\quad x\gg1\,,\\
s^{\rm TM}(x)&\sim&\frac{15\zeta(5)}{2\pi^4}\frac1{x^4}
+\frac{3\zeta(3)}{2\pi^2}\frac1{x^2}\,,\quad x\ll1\, .
\eea
\end{subequations}
 The latter  overwhelms the former (except for $\lambda_0/T\gg1$, to which
case we will return below), so the electromagnetic
self-entropy is positive, and would likely overwhelm the negative CP
energy of an atom interacting with the plate.  See Fig.~\ref{pentropy}.

In the strong coupling limit, the TE and TM self-entropies cancel.  However,
we must also consider the self-entropies of the nanoparticle.
These are easily computed from the free energies calculated from
 the potentials given in Eq.~(\ref{tatom}):
\be
F_\alpha=-\frac{T}2\Tr \mathbf{V}_n^E \bm{\Gamma}_0,\quad
F_\beta=-\frac{T}2\Tr \mathbf{V}_n^M \bm{\Gamma}_0.
\ee
Again, we regulate this with a spatial cutoff $\delta$.  An elementary
calculation yields, under the assumption that $\alpha$ posseses no
frequency dependence (consistent with the above calculations of
the interaction entropies),
\be
F_\alpha=\frac{2\alpha}{\pi\delta^4}-\frac{2\alpha}{15}\pi^3 T^4,
\ee
which diverges as $\delta\to 0$, but gives a finite entropy.
When the corresponding $S_\beta$ contribution is added, the entropy
for a perfectly conducting sphere is
\be
S=\frac{4\alpha}{15}(\pi T)^3\to\frac4{15}(\pi aT)^3.
\ee
This result agrees with that found by Balian and Duplantier \cite{balian}.
This term precisely cancels the extreme negative value for the interaction
entropy between an perfectly conducting nanosphere and a perfectly
conducting plate seen in Eq.~(\ref{mininent}).  So the total 
entropy is positive, at least in this interesting instance.

\section{Conclusions}
In this paper we have discussed situations in which
negative Casimir-Polder  entropy can arise from purely geometrical 
configurations, say between an atom and a plate, or between two atoms.
Here the atoms are characterized by static electric and magnetic 
polarizabilities.  We study this effect in this regime because
the dominant negative entropy phenomena seem to occur in
situations where the  dipole approximation is valid.
Negative entropy arises because of an interplay between E and H modes, 
or anisotropy in the polarizability of the atoms.
Sometimes the effect is perturbative, in that
negative entropy  occurs for the lowest temperatures,
but in other cases it is nonperturbative, in that the entropy starts off
positive, but changes sign for larger temperatures.

The interpretation of negative entropy is not altogether clear.
Therefore, we have examined the self-entropy of the plates and of the
nanoparticles.  In this way we have seen, at least in some rather realistic
situations, that the total entropy, including the finite self-entropies of
the bodies, may always be positive.  This argument requires no appeal to
the positive entropy of empty space.

It would be extremely interesting if such effects could be observed in
laboratory experiments.  The entropy is not such a directly observable
quantity, however, so it might prove more feasible to look at regions where
the specific heat turns negative.  This is a related phenomenon, since
\be
C_v=T\frac{\partial S}{\partial T}\, .
\ee
For a discussion of negative specific heats in the context of dissipative
  quantum systems see, e.g., Ref.~\cite{spreng} and references therein, 
as well as  Ref.~\cite{ingold09} 
for a discussion in relation to the Casimir effect.

Finally, we remark that negative entropy is not unrelated to Casimir
repulsion (see, for example, Ref.~\cite{rep}).  In both cases what is involved
is nonmonotonicity in the quantum free energy.  Here, too, experiments
revealing such phenomena in geometric contexts would be welcome.
\begin{acknowledgements}
We thank CNRS, the Simons Foundation, and the Julian Schwinger Foundation,
for partial support of this research.
\end{acknowledgements}


\begin{thebibliography}{00}
  
  \bibitem{bs}%
  \textsc{M.~Bostr\"om} and
  \textsc{Bo E.~Sernelius},
  \jr{Phys. Rev. Lett.} \textbf{84}, 4757 (2000).
  
 \bibitem{decca}%
  \textsc{R.\,S.~Decca},  
  \textsc{D.~L\'opez}, \textsc{E.~Fishbach}, \textsc{G. L. Klimchitskaya},
\textsc{D.\,E.~Krause}, and \textsc{V.\,M.~Mostepanenko}, 
  \jr{Ann. Phys. (N.Y.)}   \textbf{318}, 37 (2005).

\bibitem{mohideen}
\textsc{A.\,A.~Banishev}, \textsc{G.\,L.~Klimchitskaya}, 
\textsc{V.\,M.~Mostepanenko}, and {U.~Mohideen}, \jr{Phys. Rev. B} 
\textbf{88}, 155410 (2013).
  
\bibitem{bimonte-decca}
\textsc{G.~Bimonte}, \textsc{D. L\'opez}, and \textsc{R.\,S.~Decca},
arXiv:1509.05349v2.

\bibitem{sushkov}
\textsc{A.\,O.~Sushkov}, \textsc{W.\,J.~Kim}, \textsc{D.\,A.\,R.~Dalvit},
and \textsc{S.\,K.~Lamoreaux},
\jr{Nature Physics} \textbf{7}, 230 (2011).


\othercit                                                                    
\bibitem{dalvit} 
\textsc{D.\,A.\,R.~Dalvit},             
\textsc{P.~Milonni}, \textsc{R.~Roberts}, and 
\textsc{F.~da Rosa} (eds.),                                              
  Casimir Physics  (Springer, Berlin, 2013).



\bibitem{njp}
\textsc{I.~Brevik}, \textsc{S.\,A.~Ellingsen}, and \textsc{K.\,A.~Milton},
\jr{New J. Phys.} \textbf{8}, 236 (2006).

\bibitem{ingold}
\textsc{S.~Umrath}, \textsc{M.~Hartmann}, \textsc{G.-L.~Ingold}, and
{P.\,A.~Maia Neto},
\jr{Phys. Rev. E} \textbf{92}, 042125 (2015).

\bibitem{canaguier}
\textsc{A.~Canaguier-Durand}, \textsc{P.\,A.~Maia Neto}, \textsc{A.~Lambrecht},
and \textsc{S.~Reynaud},
\jr{Phys. Rev. Lett.} \textbf{104}, 040403 (2010).

\bibitem{ingold1}
\textsc{G.-L.~Ingold}, 
\textsc{S.~Umrath}, \textsc{M.~Hartmann}, 
\textsc{R.~Gu\'erout}, 
\textsc{A.~Lambrecht}, \textsc{S.~Reynaud}, and 
\textsc{K.\,A.~Milton}, 
\jr{Phys. Rev. E} \textbf{91}, 033203 (2015).

\bibitem{milton}
\textsc{K.\,A.~Milton},
\textsc{R.~Guer\'out}, 
\textsc{G.-L.~Ingold}, 
\textsc{A.~Lambrecht}, and \textsc{S.~Reynaud}, 
\jr{J. Phys. Condens. Matter} \textbf{27}, 214003 (2015).

\bibitem{vac}
\textsc{K.\,A.~Milton}, \textsc{P.~Parashar}, \textsc{J.~Wagner}, and
\textsc{I.~Cavero-Pel\'aez} \jr{J. Vac. Sci. Technol. B} \textbf{28} C4A8--16
(2010). 

\bibitem{cp}
\textsc{H.\,B.\,G.~Casimir} and \textsc{D.~Polder}, 
\jr{Phys. Rev.} \textbf{73}, 360 (1948).

\bibitem{bezerra}
\textsc{V.\,B.~Bezerra}, \textsc{G.\,L.~Klimchitskaya}, 
\textsc{V.\,M.~Mostepanenko}, and {C.~Romero}, \jr{Phys. Rev. A} \textbf{78},
042901 (2008).

\bibitem{bimonte}
\textsc{G.~Bimonte}, \textsc{G.\,L.~Klimchitskaya}, 
and \textsc{V.\,M.~Mostepanenko},  
\jr{Phys. Rev. A} \textbf{79}, 042906 (2009).


\bibitem{li}
\textsc{Li Yang}, \textsc{K.\,A.~Milton}, \textsc{P.~Parashar}, and
{P.~Kaluni}, manuscript in preparation.

\bibitem{shajesh} \textsc{P.~Parashar}, \textsc{K.\,A.~Milton}, 
\textsc{K.\,V.~Shajesh}, and \textsc{M.~Schaden},
\jr{Phys. Rev. D} {86}, 085021 (2012).

\bibitem{balian} \textsc{R.~Balian} and \textsc{B.~Duplantier},
\jr{Ann. Phys. (N.Y.)} \textbf{112}, 165 (1978).

\bibitem{spreng}
\textsc{B.~Spreng}, \textsc{G.-L.~Ingold}, and \textsc{U.~Weiss},
\jr{Phys. Scr. T} \textbf{165}, 014028 (2015).

\bibitem{ingold09}
\textsc{G.-L.~Ingold}, \textsc{A.~Lambrecht}, and \textsc{S.~Reynaud},
\jr{Phys. Rev. E} \textbf{80}, 041113 (2009).

\bibitem{rep}
\textsc{K.\,A.~Milton}, \textsc{E.\,K.~Abalo}, \textsc{P.~Parashar}, 
\textsc{N.~Pourtolami}, \textsc{I.~Brevik}, \textsc{S.\,\AA.~Ellingsen}, 
\textsc{S.\,Y.~Buhmann}, and \textsc{S.~Scheel}, 
\jr{Phys. Rev. A} \textbf{91}, 042510 (2015).

  
  
  
  
\end{thebibliography}
\end{document}